\documentclass[useAMS,usenatbib,usegraphicx]{mn2e}
\usepackage{epsfig}
\usepackage{amsmath} 
\usepackage{rotating}           
\usepackage{color}     
\usepackage{graphicx}
\usepackage{times}
\usepackage{upgreek} 

\def\kms{km ${\rm s}^{-1}$}

\def\Mo{M$_\odot$}

\def\ccm {$\hbox{{\rm cm}}^{-3}$}    
\def\scm  {$\hbox{{\rm cm}}^{-2}$}    
\def \AL {$\alpha $}     
\def \HI {H{\sc \,i}}
\def \HII {H{\sc \,ii}}

\def\lapp{\ifmmode\stackrel{<}{_{\sim}}\else$\stackrel{<}{_{\sim}}$\fi}
\def\gapp{\ifmmode\stackrel{>}{_{\sim}}\else$\stackrel{>}{_{\sim}}$\fi}
\title[Cold neutral gas and the star formation history]{The evolution of cold neutral gas and the star formation history} 
\author[S. J. Curran et al.]{S. J. Curran\thanks{Stephen.Curran@vuw.ac.nz}\\
School of Chemical and Physical Sciences, Victoria University of Wellington, PO Box 600, Wellington 6140, New Zealand\\
}
\begin{document}

 \date{Accepted ---. Received ---; in original form ---}

\pagerange{\pageref{firstpage}--\pageref{lastpage}} \pubyear{2019}

\maketitle

\label{firstpage}
\begin{abstract}
  There is a well known disparity between the evolution the star formation rate density, $\psi_{*}$, and the abundance
  of neutral hydrogen (\HI), the raw material for star formation. Recently, however, we have shown that $\psi_{*}$ may
  be correlated with the fraction of {\em cool} atomic gas, as traced through the 21-cm absorption of \HI. This is
  expected since star formation requires cold ($T\sim10$~K) gas and so this could address the issue of why the star
  formation rate density does not trace the bulk atomic gas. The data are, however, limited to redshifts of $z\lapp2$,
  where both $\psi_{*}$ and the cold gas fraction exhibit a similar steep climb from the present day ($z=0$), and so it
  is unknown whether the cold gas fraction follows the same decline as $\psi_{*}$ at higher redshift.  In order to
  address this, we have used unpublished archival observations of 21-cm absorption in high redshift damped
  Lyman-$\alpha$ absorption systems to increase the sample at $z\gapp2$. The data suggest that the cold gas fraction
  does exhibit a decrease, although this is significantly steeper than $\psi_{*}$ at $z\sim3$. This is, however,
  degenerate with the extents of the absorbing galaxy and the background continuum emission and upon removing these, via
  canonical evolution models, we find the mean spin temperature of the gas to be $\left<T_{\rm spin}\right>
  \approx3000$~K, compared to the $\approx2000$~K expected from the $\psi_{*}T_{\rm spin}\approx100$
  ~\Mo~yr$^{-1}$~Mpc$^{-3}$~K fit at $z\lapp2$. These temperatures are consistent with the observed high neutral
  hydrogen column densities, which require $T\lapp4000$~K in order for the gas not to be highly ionised.  
\end{abstract} 
 

\section{Introduction} 
\label{intro}

Galaxies intervening the sight-lines to more distant Quasi-Stellar Objects (QSOs) allow study of the neutral gas in the
distant Universe, through the absorption of the background continuum radiation by the Lyman-$\alpha$ ($\lambda
=1215.67$~\AA) transition of hydrogen. These so-called {\em damped Lyman-$\alpha$ absorbers} (DLAs),
where the neutral hydrogen column density exceeds $N_{\rm HI}\geq2\times10^{20}$ atoms \scm, may contain up to 80\% of
the neutral gas mass density in the Universe \citep{phw05}. The detection of this transition is restricted to redshifts
of $z\gapp1.7$ (the first four billion years of the Universe's history) by ground-bases telescopes, where this
ultra-violet band transition is redshifted into the atmospheric window at visible wavelengths. Apart from space-based
observations (e.g. \citealt{rtsm17}), hydrogen at lower redshifts can be detected through the spin-flip
($\lambda =21.1$ cm) transition, which occurs in the radio band.
 
\begin{figure}
\centering \includegraphics[angle=-90,scale=0.48]{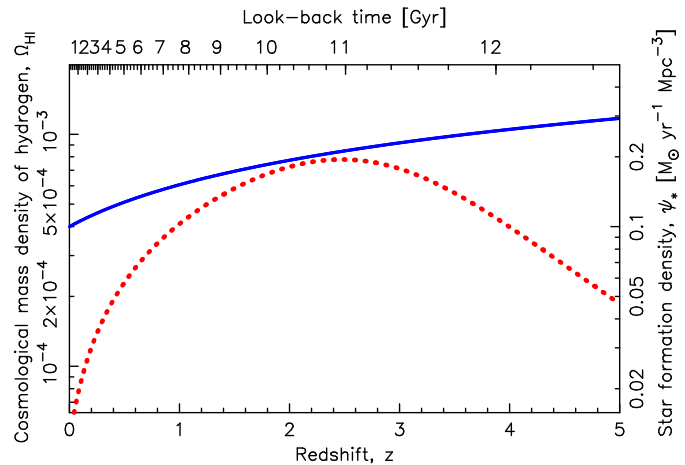}
\caption{The cosmological mass density of neutral hydrogen (solid trace \& left scale, \citealt{cmp+17}) and the star
  formation density (dotted trace \& right scale, \citealt{hb06}).} 
\label{omega-SFR}
\end{figure}

Observations of both \HI\ 21-cm emission  \citep{zvb+05,lcb+07,bra12,dsmb13,rzb+13,hsf+15} and Lyman-$\alpha$ absorption 
\citep{hb06,bbg+13,bwc13,ssb+13,md14,zjd+14} show that there is little evolution in the mass density of
neutral hydrogen, which is in stark contrast to the steep evolution in the cosmic star formation rate
(Fig.~\ref{omega-SFR}).
However,
both Lyman-$\alpha$ absorption and 21-cm emission trace all of the neutral gas ($T\lapp10\,000$~K), whereas only
the gas clouds which are cool enough ($T\sim10$~K) to collapse under their own gravity can initiate star formation. This
cool component of the gas is detected through the {\em absorption} of the 21-cm radiation from a
background radio-loud QSO (quasar). 
If the Lyman-\AL\ and 21-cm absorption trace the same sight-line, comparison of the 21-cm absorption strength with the
total neutral hydrogen column density can, in principle, yield the spin temperature of the gas, $T_{\rm spin}$, although
this is degenerate with the fraction of the background radio flux intercepted by the absorbing gas. 

Although this {\em covering factor} is difficult to quantify, by accounting for the angular diameter distances to the
absorbing galaxy and background continuum source, we can at least remove any bias introduced by an expanding Universe from
the flux coverage. This yields the spin temperature degenerate with the ratio of the projected sizes of the absorber and
the continuum source, $T_{\rm spin} (d_{\rm QSO}/d_{\rm abs})^2$. Using this method, we \citep{cur17,cur17a} have shown
that the reciprocal of this, $(1/T_{\rm spin}) (d_{\rm abs}/d_{\rm QSO})^2$, increases by a similar amount as the star
formation rate from $z_{\rm abs}=0$ to the peak of star formation at $z_{\rm abs}\sim2$. That is, for a non-evolving
$d_{\rm abs}/d_{\rm QSO}$ ratio, the spin temperature is anti-correlated with $\psi_{*}$, as would be expected upon the
basis that stars can only form out of the coldest gas.  Being limited to $z_{\rm abs}\lapp2$, however, means that we do
not know whether $(1/T_{\rm spin}) (d_{\rm abs}/d_{\rm QSO})^2$ exhibits a downturn at higher redshift, thus truly
tracing the star formation history, rather than a coincidental similar factor of increase over $0\lapp z_{\rm abs}\lapp2$.
 In order to address this, here we add a dozen previously unpublished high redshift searches for 21-cm
absorption in DLAs from the data archive of the Giant Metrewave Radio Telescope (GMRT).

\section{Data acquisition and reduction}

\begin{figure*} 
\vspace{10.5cm}  
\includegraphics{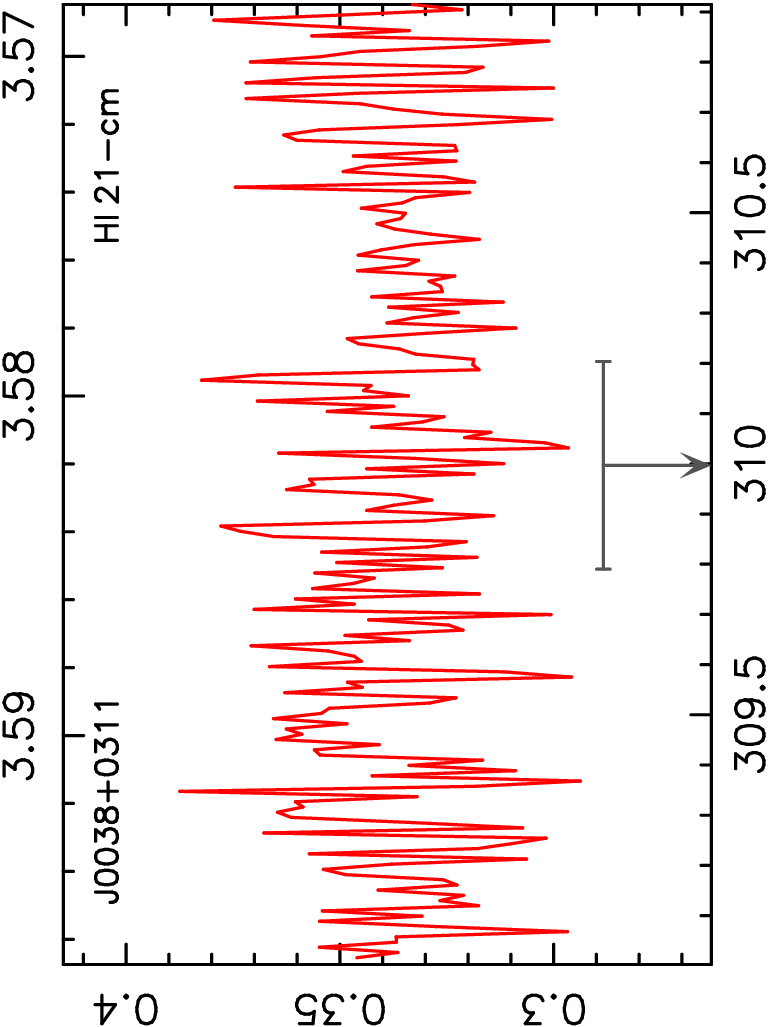}  
\includegraphics{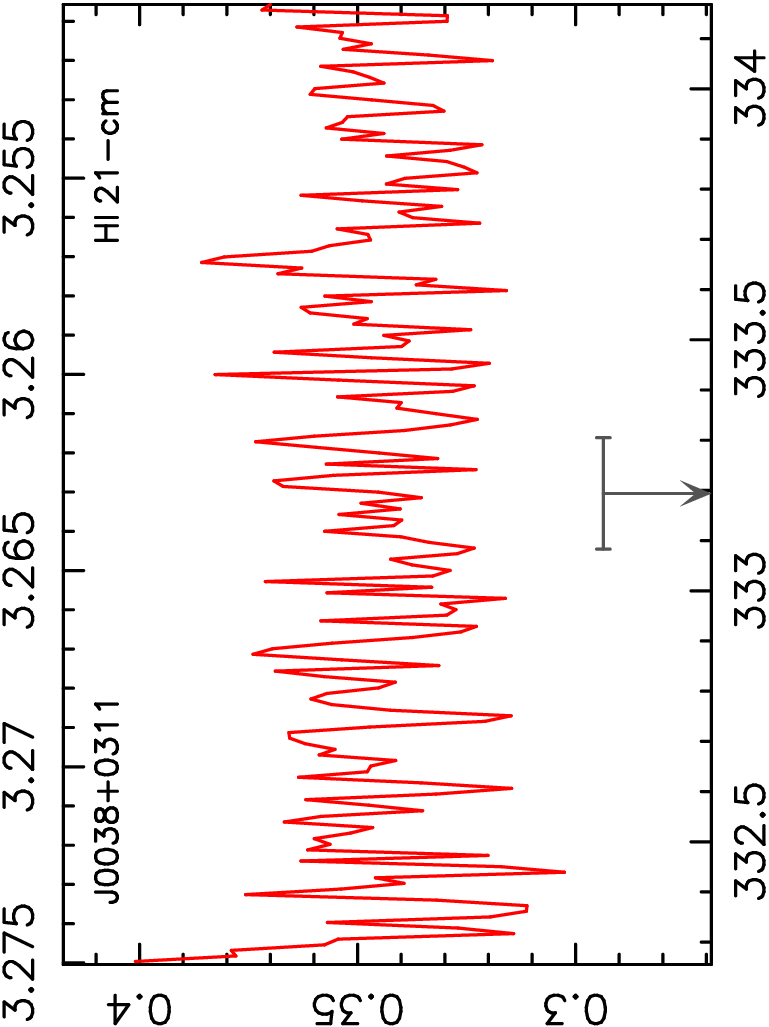}  
\includegraphics{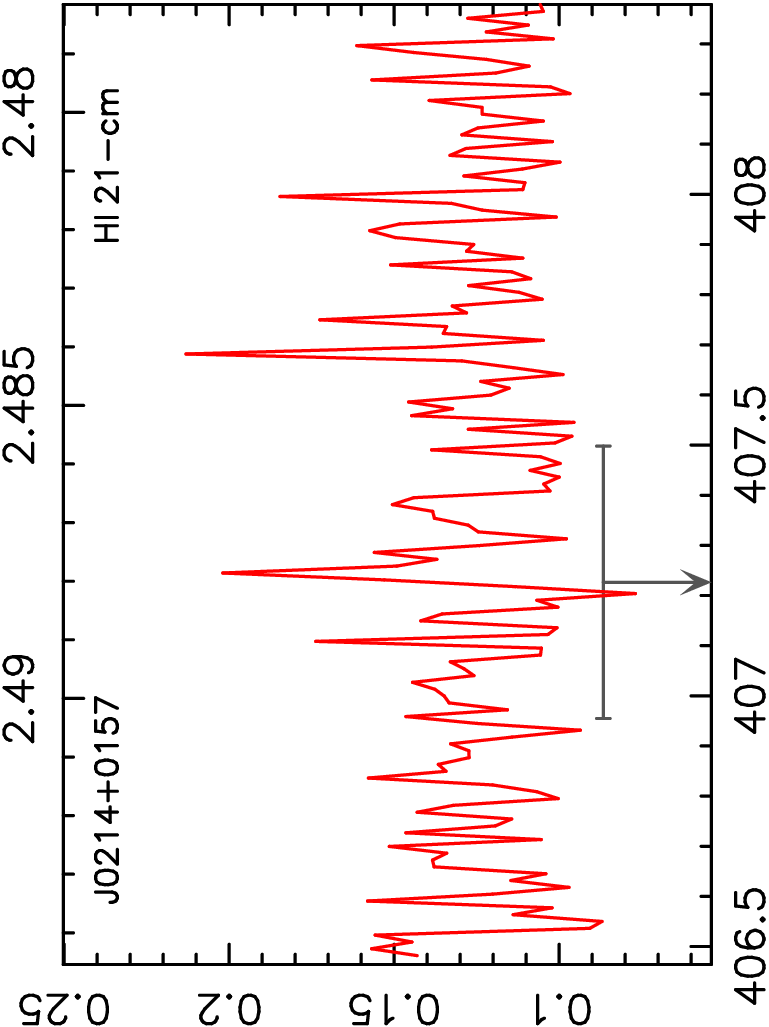} 
\includegraphics{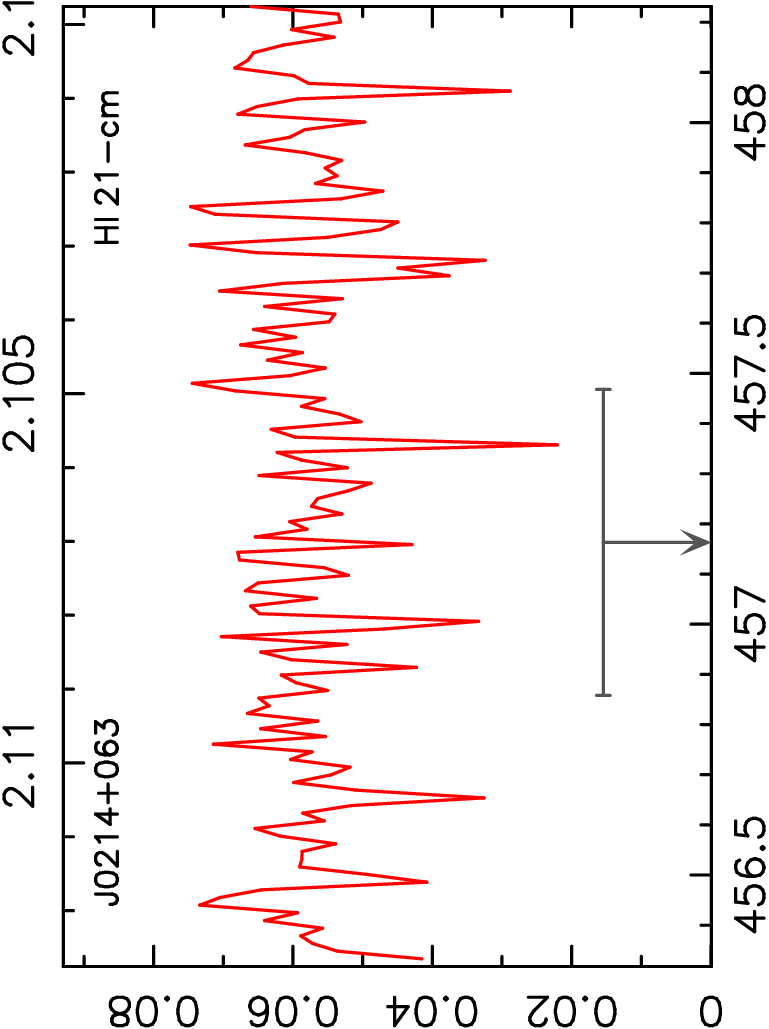}
\includegraphics{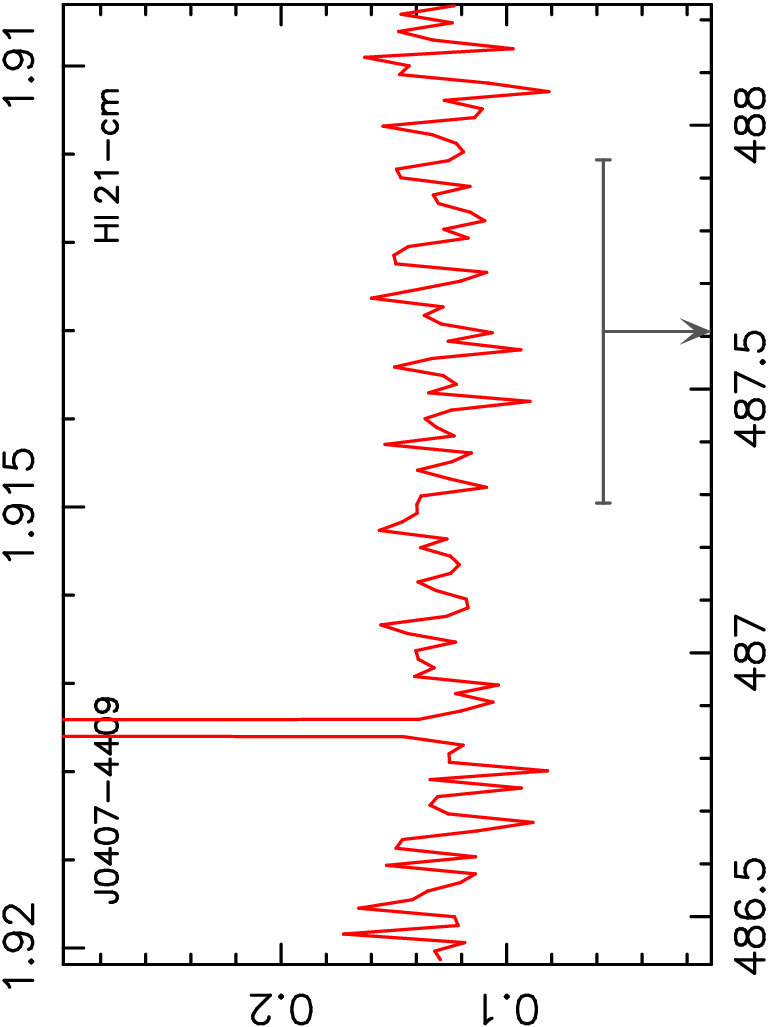} 
\includegraphics{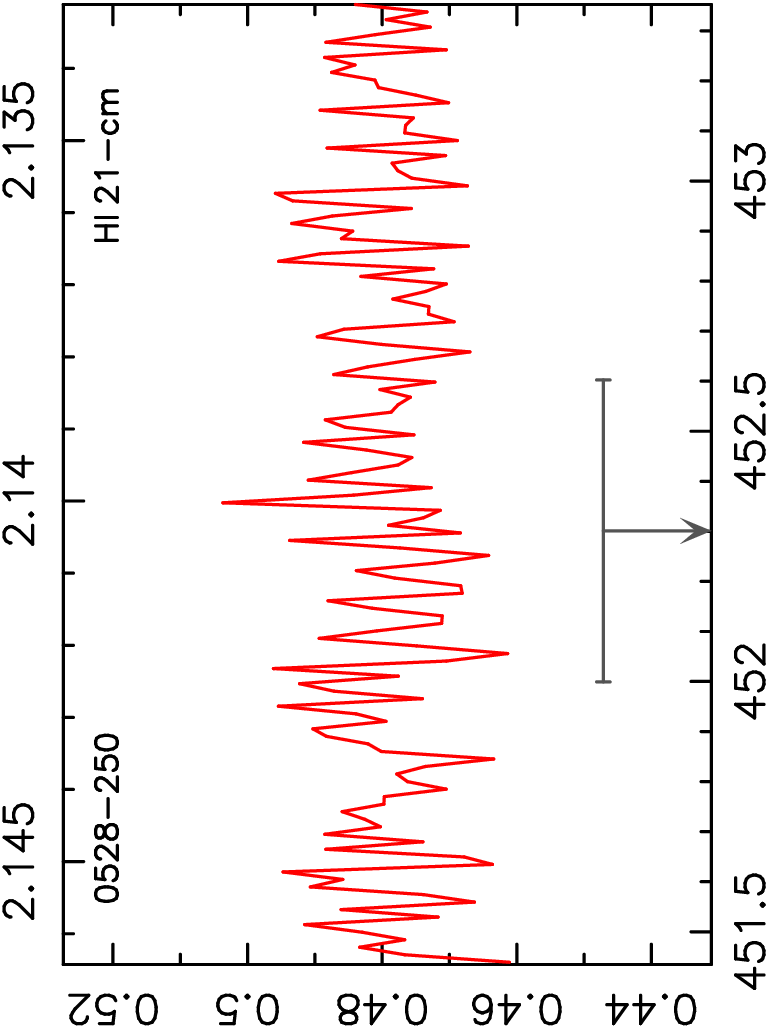}  
\includegraphics{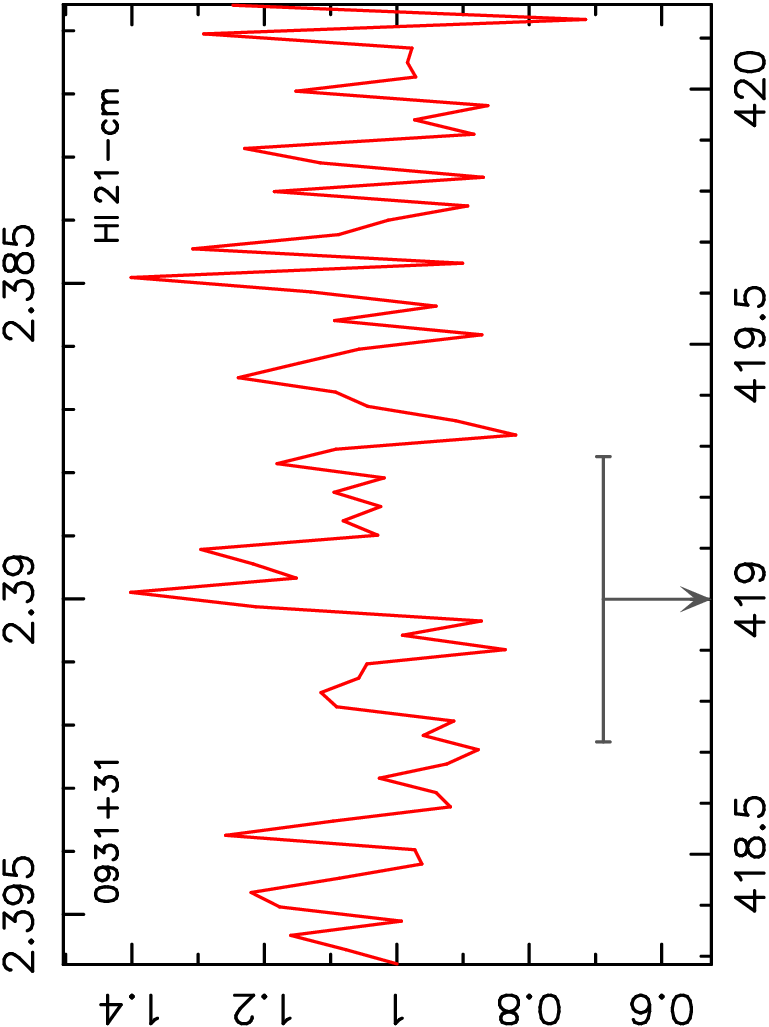} 
\includegraphics{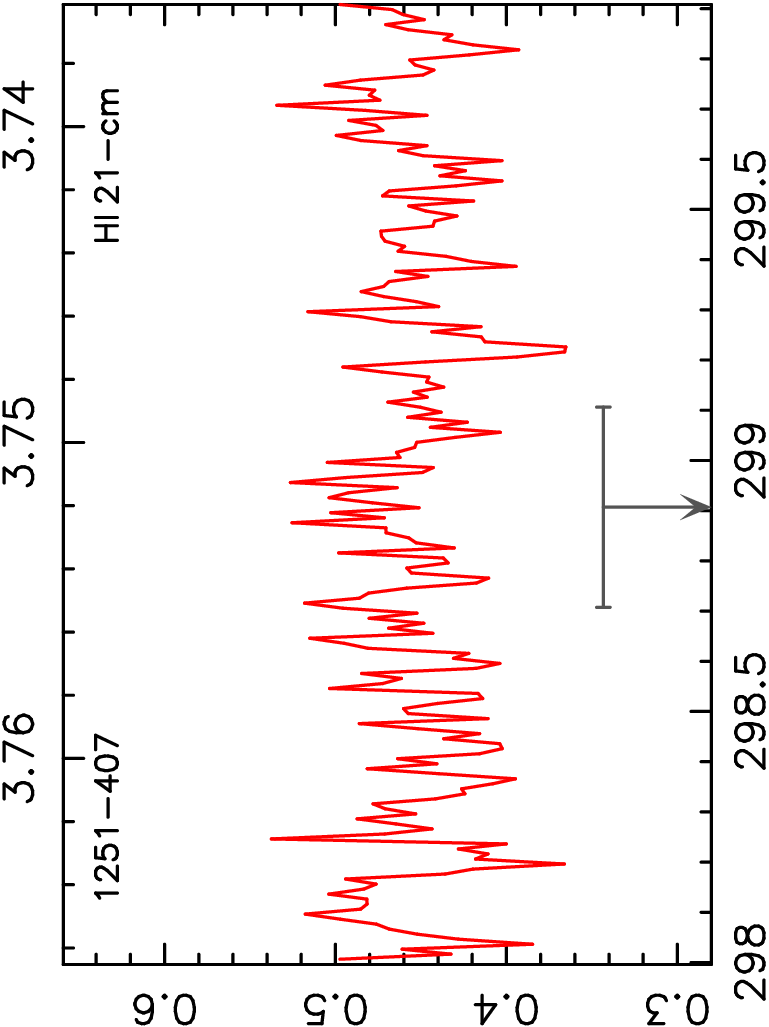} 
\includegraphics{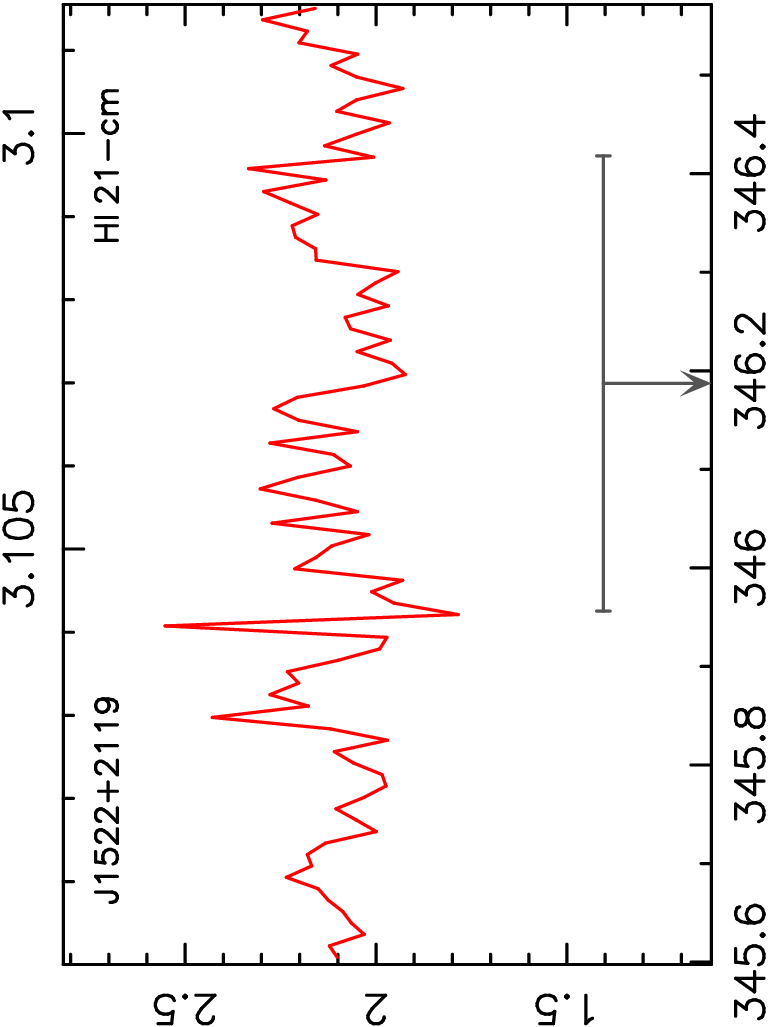}  
\includegraphics{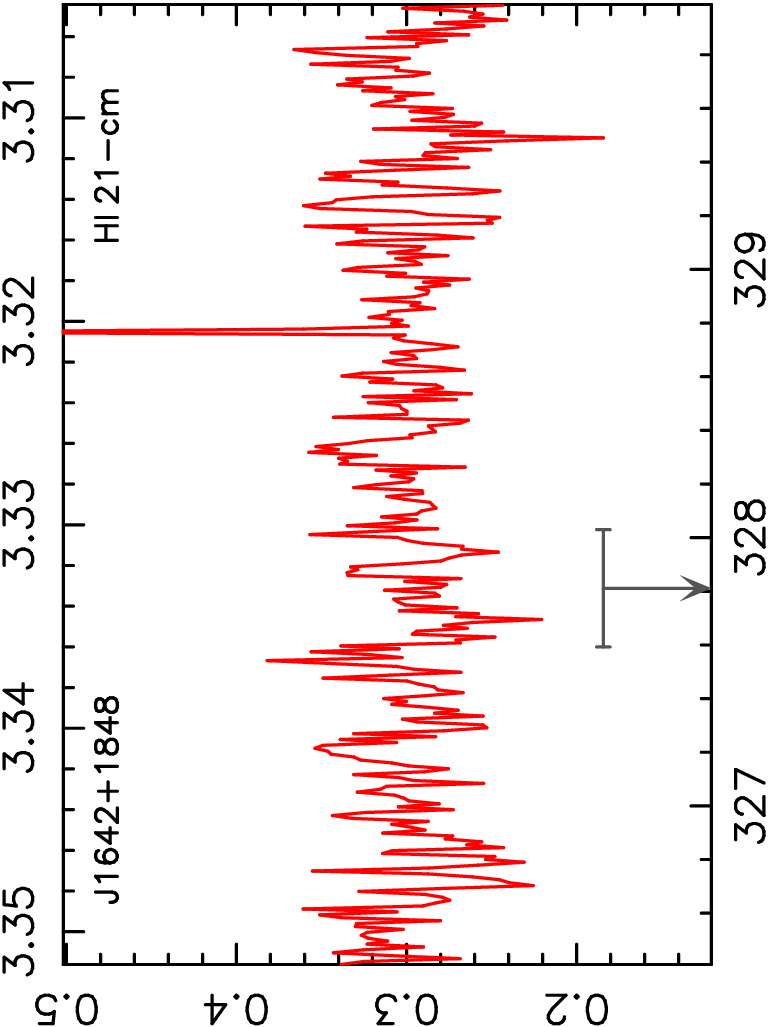} 
\includegraphics{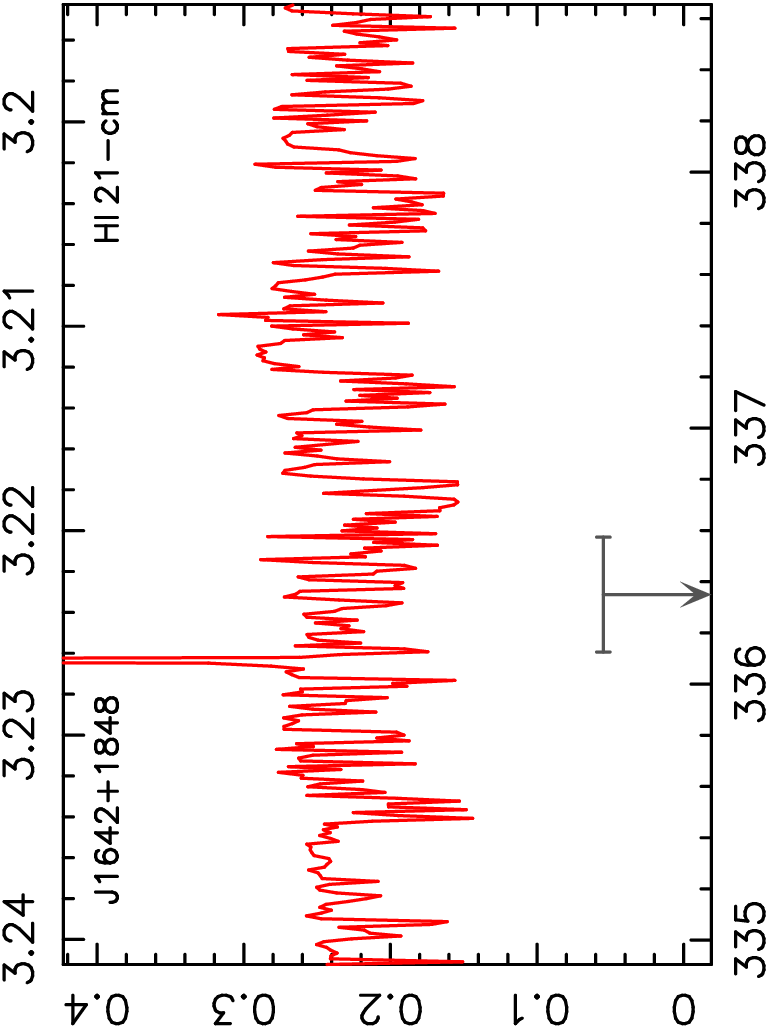} 
\includegraphics{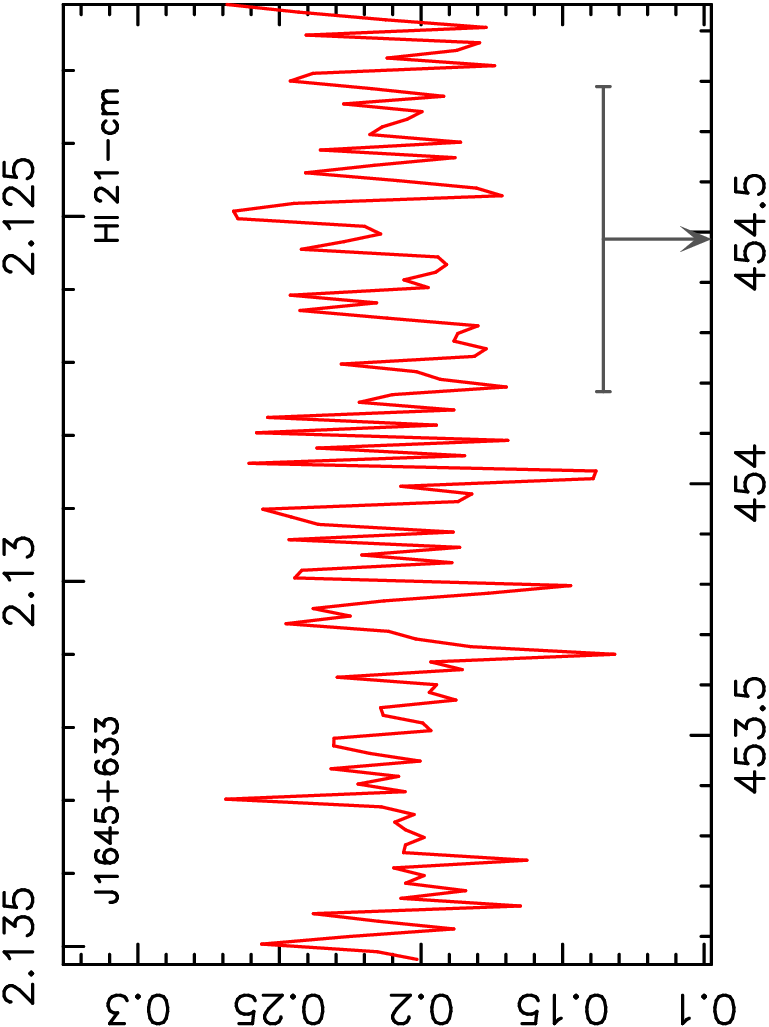} 
\caption{The reduced spectra shown at a spectral resolution of 10 \kms.  The ordinate gives the flux density [Jy] and the
  abscissa the barycentric frequency [MHz]. The scale along the top shows the redshift of \HI\ 21-cm over the frequency
  range and the downwards arrow shows the expected frequency of the absorption from the optical redshift, with the
  horizontal bar showing a span of $\pm200$ \kms\ for guidance.}
\label{spectra}
\end{figure*}

The GMRT is the longest serving large interferometer capable of observing at the required frequencies ($\lapp470$~MHz
for 21-cm at $z\gapp2$), thus having the most comprehensive archive, in addition to being able to reach the required
sensitivities \citep{cur18}.  Radio frequency interference (RFI) meant that many of the sources in the archive could not
be reduced and some flagging of badly affected data were required on the dozen remaining sight-lines (Table~\ref{obs}).
These had all been observed using the full 30 antenna array, with 3C\,48, 3C\,147 and 3C\,298 used for bandpass
calibration and a nearby bright, unresolved radio source for phase calibration. Upon downloading the data, these
were calibrated and flagged with the {\sc miriad} interferometry reduction package.
\begin{table*}
\centering
 \caption{The unpublished archival data not completely corrupted by RFI. The name of the background continuum source is 
followed
    by its redshift. The redshift and neutral hydrogen column density of the absorber are then given, followed by the
    observed frequency, GMRT project number, the date of the observation and the total observing time, $t_{\rm
      obs}$. $\Delta S$  is the rms noise reached per 10 \kms\ channel, $S_{\rm obs}$ is the observed flux
    density and $\tau_{3\sigma}$ the derived optical depth limit, where $\tau_{3\sigma}=-\ln(1-3\Delta S/S_{\rm meas})$ is
    quoted for these non-detections.}
\begin{tabular}{@{}l c c  c  c c  r c c c c  @{}} 
\hline
\smallskip
QSO & $z_{\rm QSO}$ & $z_{\rm abs}$ & $N_{\text{\HI}}$ [\scm] & $\nu_{\rm obs}$ [MHz] & Proj. & Date & $t_{\rm obs}$ [h]& $\Delta S$  [mJy] & $S_{\rm obs}$ [Jy] & $\tau_{3\sigma}$ \\
\hline
SDSS\,J003843.98+031120.8  &	3.674 & 3.582   & 21.60  & 310.00 & 24\_025  & 1/9/2013 & 6.17 & 20.1 & 0.338 & $<0.178$\\
... & ...                                                           & 3.263  &   20.29&  333.19 & 24\_025 & 9/8/2013 & 4.41 & 17.6 & 0.345  &   $<0.153$\\
SDSS\,J021435.77+015702.8 & 3.285     & 2.488	 &  20.88 & 407.23 & 30\_068 & 2/9/2016 & 1.53  & 21.8 &   0.125 &    $<0.741$\\
SDSS\,J021437.02+063251.3  & 2.311 &     2.107     &20.80 & 457.16 & 30\_068  & 25/8/2016 & 4.07  & 8.87 &   0.058  &    $<0.614$\\
2MASS\,J04071807--4410141     & 3.020     & 1.9130      &   20.60  & 487.61 & 30\_068 & 2/9/2016  & 1.97   & 20.1 &  0.134&    $<0.597$\\	
\protect[HB89]\,0528--250& 2.813&  2.1404	&21.00 & 452.30&  30\_068	& 5/9/2016     & 2.00    & 8.53 & 0.479 &   $<0.053$	\\			      		
B2\,0931+31 &  2.895     &   2.390 &--- &    419.00&  30\_068 &   1/10/2016 &   2.31   &  114 &     1.050    &  $<0.390$\\ 
  PKS\,1251--407             & 4.460& 3.752 & 20.30  & 298.91  & 24\_055 & 16/9/2013 &  3.98  & 33.4 & 0.454 & $<0.221$\\
SDSS\,J152219.67+211957.3 & 3.225  & 3.103       &  20.55      & 346.19     &   24\_055 &  14/5/2013  & 6.59    &   129 &  2.100 &  $< 0.184$\\                                             	
SDSS\,J164208.62+184859.4   &  3.333      & 3.150     &  20.68    &342.27   & 24\_055  & 12/7/2013     &   3.52   & 30.7 &   0.299   &  $<0.368$\\

...                                                  &  ...              & 3.223     &  20.51    & 336.35   & 24\_055	& 13/7/2013   & 6.25    & 22.5   &   0.252    &   $< 0.311$\\
WISE\,J164558.54+633010.8 & 2.379	  & 2.1253	&20.55 & 454.49 & 30\_068	   & 25/8/2016 &   1.17   &  25.7  &  0.210    &   $< 0.453$	\\  
\hline
\end{tabular}
\label{obs}  
\end{table*} 
For each of these sources, after calibration, the two polarisations were averaged and a cube produced, from
which a spectrum was extracted (Fig.~\ref{spectra}).
The addition of these absorbers, takes the total number to 85, cf. the 74 previously \citep{cur17}.\footnote{From 11
  additional sight-lines.  No value for the column density could be found for the $z_{\rm abs}= 2.390$ absorber towards
  B2\,0931+31.}  All of the new absorbers  have redshifts of $z_{\rm abs}\geq1.913$, where there are currently 36 DLAs with
published searches for 21-cm absorption.\footnote{These have been compiled from  \citet{dm78,bs79,bw83,ck00,kcsp01,kgc01,kpec09,kem+12,kps+14,bdv01,kc01a,kc02,cmp+03,ctp+07,ctd+09,ykep07,gsp+09,gsp+09a,ekp+12,sgp+12,rmgn13,kan14}.}

\section{Analysis}
\subsection{Spin temperature--covering factor degeneracy}
\label{stcfg}

The total neutral atomic hydrogen column density, $N_{\rm HI}$ [\scm], is related to the velocity integrated optical
depth of the \HI\ 21-cm absorption [\kms] via \citep{wb75}
\begin{equation}
N_{\rm HI}  =1.823\times10^{18}\,T_{\rm  spin}\int\!\tau\,dv,
\label{enew_full}
\end{equation}
where $T_{\rm spin}$ [K] is the harmonic mean spin temperature --  the density weighted average of the
spin temperature of the absorbing gas along the sight-line.
This is a measure of the population of the lower hyperfine level ($F=1$),
where the gas can absorb 21-cm photons \citep{pf56}, relative to the upper hyperfine level ($F=2$).  
However, we cannot
measure $\int\!\tau\,dv$ directly, since the observed optical depth, which is the ratio of the line depth, $\Delta S$,
to the observed background flux, $S_{\rm obs}$, is related to the intrinsic optical depth via
\begin{equation}
\tau \equiv-\ln\left(1-\frac{\tau_{\rm obs}}{f}\right) \approx  \frac{\tau_{\rm obs}}{f}, {\rm ~for~}  \tau_{\rm obs}\equiv\frac{\Delta S}{S_{\rm obs}}\lapp0.3,
\label{tau_obs}
\end{equation}
where the covering factor, $f$, is the fraction of $S_{\rm obs}$ intercepted by the absorber.  In the optically thin
regime (where $\tau_{\rm obs}\lapp0.3$)\footnote{This applies to all of the DLAs detected in 21-cm absorption, where the
  observed optical depths span $\tau_{\rm obs}=0.05-0.26$.}, 
 Equ.~\ref{enew_full} can be
approximated as
\begin{equation}
N_{\text{\HI}}  \approx 1.823\times10^{18}\,\frac{T_{\rm  spin}}{f}\int\!\tau_{\rm obs}\,dv.
\label{enew}
\end{equation}
That is, comparison of the 21-cm line strength with the total column density  yields the spin temperature
degenerate with the covering factor ($T_{\rm spin}/f$).

In order to obtain the physically interesting spin temperature the covering factor must be known.  This requires
knowledge of the absorbing cross-section, the extent of the background continuum, as well as the alignment between the
absorber and continuum source (see \citealt{cur17}) and so is generally unknown.  Estimates of the covering factor
assume that this is equal to the ratio of the compact unresolved component's flux to the total radio flux
(e.g. \citealt{bw83,kps+14}). However, this gives no information on the depth of the absorption when the extended
continuum emission is resolved out nor can it yield information on the absorber and how effectively it covers the
emission. This requires high resolution, highly sensitive observations of the absorption across the emission region
(e.g. \citealt{lbs00}) at low frequencies, thus requiring the Square Kilometre Array (SKA): Phase-1 will be an order of
magnitude more sensitive than the GMRT at $z\gapp3$, with phase-2 increasing the sensitivity by another order of
magnitude \citep{cur18}.

Until such high resolution imaging of the absorption becomes available at the required frequencies, we can use a
statistical approach:  
In the small
angle approximation, the covering factor can be obtained from 
\begin{equation}
f = \left\{   
\begin{array}{l l}
\left(\frac{d_{\rm abs} DA_{\rm QSO}}{d_{\rm QSO} DA_{ \rm abs}}\right)^2 & \text{ if } \theta_{\rm abs} < \theta_{\rm QSO}\\
  1  & \text{ if } \theta_{\rm abs} \geq\theta_{\rm QSO},\\
\end{array}
\right.  
\label{f}
\end{equation}
\citep{cur12}, where the angular diameter distance to a source is 
\begin{equation}
DA = \frac{DC}{z+1},{\rm ~where~} 
\label{equ:DA}
\end{equation}
\[
DC = \frac{c}{H_0}\int_{0}^{z}\frac{dz}{\sqrt{\Omega_{\rm m}\,(z+1)^3 +  (1-\Omega_{\rm m} - \Omega_{\Lambda})\,(z+1)^2 + \Omega_{\Lambda}}}
\]
is the line-of-sight co-moving distance (e.g. \citealt{pea99}), in which $c$ is the speed of light and $H_0$ the Hubble constant. %

For a standard $\Lambda$ cosmology with $H_{0}=71$~km~s$^{-1}$~Mpc$^{-1}$, $\Omega_{\rm matter}=0.27$ and
$\Omega_{\Lambda}=0.73$, this gives a peak in the angular diameter distance at $z\approx1.6$, which has the consequence
that below this redshift both $DA_{\rm DLA} \ll DA_{\rm QSO}$ and $DA_{\rm DLA} \sim DA_{\rm QSO}$ are possible, whereas
above $z_{\rm abs}\sim1.6$, only $DA_{\rm DLA} \sim DA_{\rm QSO}$ is possible. This leads a mix of angular diameter
distance ratios ($DA_{\rm abs}/DA_{ \rm QSO}$) at low redshift but exclusively high ratios ($DA_{\rm abs}/DA_{ \rm
  QSO}\sim1$) at high redshift \citep{cw06}. Thus, there is a clear bias introduced by the geometry of the Universe, which must be
accounted for when evaluating the spin temperature. Otherwise this  leads to an apparent mix of spin
temperatures at $z\lapp1$ and exclusively high spin temperatures at $z\gapp1$ \citep{kc02}.

For covering factors of less than unity, which is generally expected to be the case (\citealt{cur17}, see also
\citealt{kps+14}), this gives the spin temperature degenerate with the ratio of the absorber--emitter size,
\[
\frac{1}{T_{\rm spin}}\left(\frac{d_{\rm abs}}{d_{\rm QSO}}\right)^2 = \frac{1.823\times10^{18}}{N_{\text{\HI}}}\left(\frac{DA_{\rm abs}}{DA_{\rm QSO}}\right)^2 \int\!\tau_{\rm obs}\,dv 
\]
\begin{equation}
~~~~~ = \frac{1.823\times10^{18}}{N_{\text{\HI}}}\left(\frac{DA_{\rm abs}}{DA_{\rm QSO}}\right)^2 \frac{\Delta S}{S_{\rm obs}}\Delta v, 
\label{equ:NoverT}
\end{equation}
where $\Delta S$ is the line depth in the case of a detection or $3\sigma_{\rm rms}$, where $\sigma_{\rm rms}$ is the r.m.s. noise, per $\Delta v$ channel for a 
non-detection.\footnote{In the literature the spectral resolutions span a
  large range of values \citep{cur17a}, and so we re-sample the r.m.s. noise levels to a common channel width, which is
  then used as full-width half maximum (FWHM) of the putative absorption profile.} 

\begin{figure}
\centering \includegraphics[angle=-90,scale=0.48]{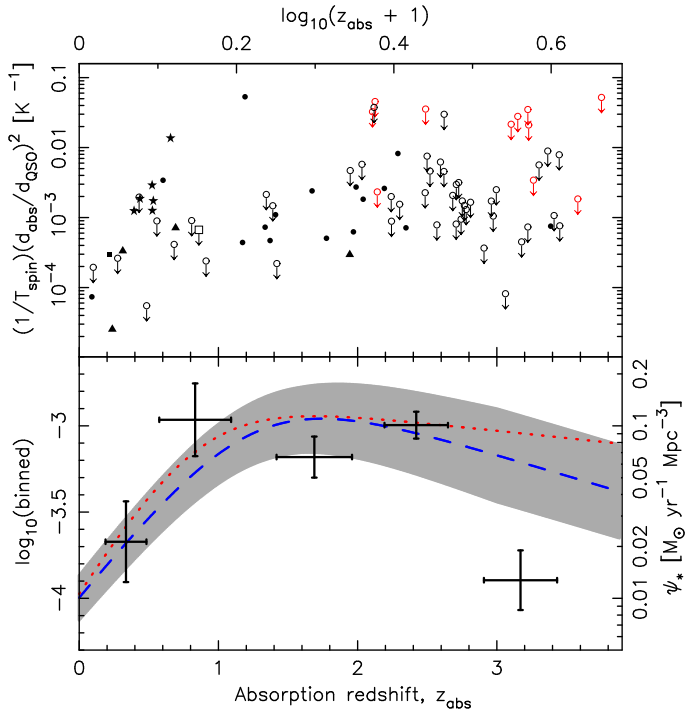}
\caption{The $(1/T_{\rm spin}) (d_{\rm abs}/d_{\rm QSO})^2$ distribution. The top panel shows the individual values,
  where the filled symbols are those detected in 21-cm absorption and the unfilled $3\sigma$ upper limits, with the
  coloured symbols representing the new data.  The shapes indicate the morphology of the absorbing galaxy: star--spiral,
  square--dwarf, triangle--low surface brightness, circle--unknown.  In the bottom panel, the error bars show
  $\pm1\sigma$ from the mean (see main text). The right hand scale shows the star formation rate density, with the
  dotted curve showing the previous fit to the evolution (Fig.~\ref{omega-SFR}) and the broken curve a more recent fit,
  where the shaded region shows the $\pm1\sigma$ uncertainty \citep{bwc13}.}
\label{spin-bcw}
\end{figure}

Adding the new data to the previous gives the distribution shown in Fig. \ref{spin-bcw}.  In the binned data, the limits
are included via the Kaplan--Meier estimator, a fundamental tenet in non-parametric  survival analysis \citep{km58}, 
which gives a  maximum-likelihood estimate based upon the parent population \citep{fn85}.
From the binning, we see that the increase in $(T_{\rm spin}) (d_{\rm QSO}/d_{\rm abs})^2$ does indeed continue with
redshift, although this is somewhat steeper than the decline of the star formation density. However, the decline in
$\psi_{*}$ has also steepened since previous estimates (\citealt{hb06} and references therein), due to newer estimates
of the dust obscuration of the high redshift ultra-violet photometry \citep{bwc13}. This is confirmed by far-infrared
photometry, which is much less attenuated by dust \citep{bbg+13}.  While the radio-band data are impervious to the
effects of intervening dust, the UV emission from the QSO will certainly be attenuated and depletion of metallic species
does indicate the presence of dust in DLAs (e.g. \citealt{lbp02}, but see also \citealt{whp06}), although the degree of
reddening suggests that this is low \citep{mb16}. Alternatively, in the binning of the data, the high redshift bin
contains only a single detection, upon which the censored data are estimated.  That is, the high redshift bin may be
better considered a lower limit and so further 21-cm detections are required in order to rule out further
corrections to the high redshift evolution of $\psi_{*}$.

\subsection{Absorber size}
\label{as}

\begin{figure}
\centering 
\includegraphics[angle=270,scale=0.48]{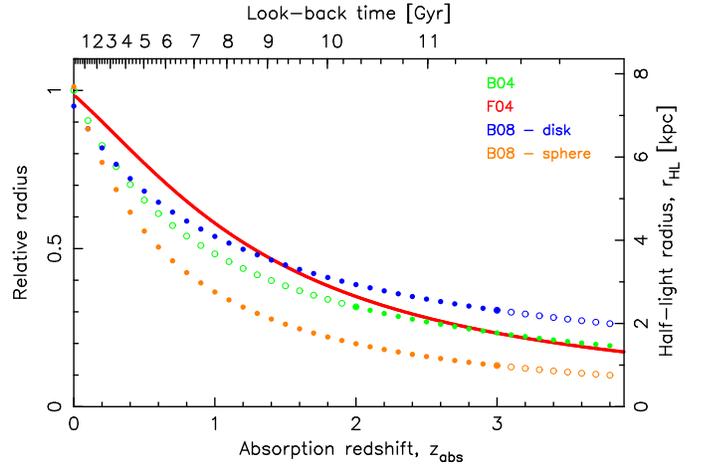}
\caption{The evolution of massive galaxy size as shown by the fits of F04 \citep{fdg+04}, B04 \citep{bib+04} at $z\geq2$
  and B08 (\citealt{btc+08}, disk-like and spheroid-like) at $z\leq3$, where the hollow circles show the extrapolation
  based upon the observed ranges (full circles). The right hand axis shows the mean half-light radius \citep{bib+04} to
  which the other curves are normalised.}
\label{abs_evolv}
\end{figure}

Asides from this last high redshift bin, the bins in Fig.~\ref{spin-bcw} exhibit some scatter around the star formation
rate density fits.  This is not unexpected, given that the mean $d_{\rm abs}/d_{\rm QSO}$ is assumed to be constant with
redshift, which may not be the case: For massive ($\geq10^{11}$ \Mo) galaxies, which are the easiest to resolve at high
redshift, there is a well documented evolution in size, where large galaxies dominate the low redshift population
\citep{bmce00} and dwarf galaxies the high redshift population \citep{lf03}, although massive DLA hosts have
been observed out to $z\sim4$ \citep{dgo98}.
Due to hierarchical build-up, one may also
expect a size evolution in DLAs and, if similar to the massive galaxies (several low redshift DLAs have been identified
as spirals), we expect a similar decrease in size with redshift. Furthermore, both disk and spheroid like
galaxies exhibit a decrease in size with redshift (Fig. \ref{abs_evolv}) and, while imaging of DLA hosts is difficult
due to the bright background QSO, at low redshift, where this is less challenging, some DLAs have been identified to be
spirals (see e.g. Fig.~\ref{spin-bcw}).

Since it is close to a mean of the other curves, in addition to providing an absolute size, we use the fit of
\citet{bib+04} to evolve the absorber sizes with redshift, i.e. $r_{\rm HL} =7.6(z_{\rm abs}+1)^{-1.05}$ kpc.  This
corresponds to $r_{\rm HL}\approx8$ kpc at $z\approx0$ for the optical extent of the galaxy, although the \HI\ disk is
generally considerably larger than this \citep{wbd+08}. For instance, DLA column densities ($N_{\rm HI}\gapp10^{20}$
\scm) have been mapped to radii of up $r\approx40$~kpc in nearby galaxies
\citep{bpb02,zmsw05,ckb08,rsa+15,rsa+16,rcm+18}, including the Milky way \citep{kk09}, as well as in distant ($z\sim1$)
galaxies (\citealt{pbk+10a,bmk+13a}).  We therefore use an \HI\ radius of $r_{\rm abs} = 5 r_{\rm HL}$ and evolve this
according to \citet{bib+04}.

\subsection{Radio source size}
\label{rss}

Quasars also exhibit a redshift evolution, although the size of the source is also dependent upon the
observed frequency:
\begin{figure}
\centering 
\includegraphics[angle=270,scale=0.48]{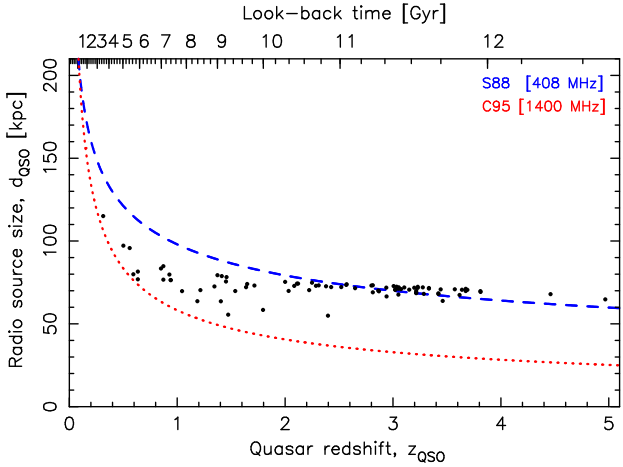}
\caption{The evolution of quasar size at 408 MHz (S88, \citealt{sin88}) and 1400 MHz (C95,\citealt{cz95}).
The circles show the values for the background sources of the sample, obtained by interpolating/extrapolating between
the curves.}
\label{quasar_evolv}
\end{figure} 
The curves in Fig.~\ref{quasar_evolv} show the evolution of quasar size at 408 and 1400~MHz.\footnote{Where we 
have converted the former \citep{sin88} from the  angular sizes using contemporary cosmological parameters (Equ.~\ref{equ:DA}).} 
Converting the power-law fits of the low \citep{sin88} and high \citep{cz95} frequency fits (Fig.~\ref{quasar_evolv}),
we obtain $d_{\rm QSO} = 97.7\,z_{\rm QSO}^{-0.307}$ at 408~MHz and $d_{\rm QSO} = 58.2\,z_{\rm QSO}^{-0.520}$ at
1400~MHz, which we interpolate between in order to determine the QSO size at the observed frequency.

\section{Results and discussion}

\subsection{Canonical values}
\label{can}

\begin{figure}
\centering 
\includegraphics[angle=270,scale=0.45]{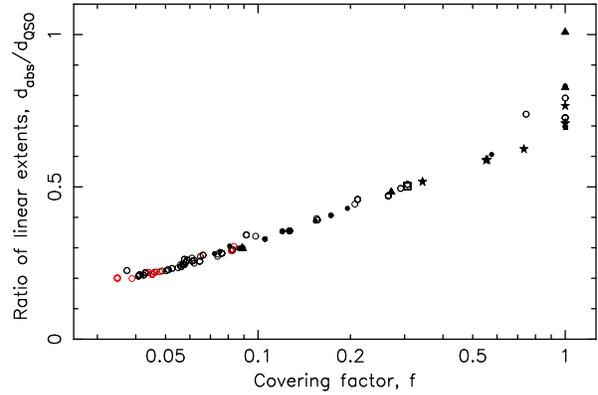}
\caption{The absorber--quasar size ratio versus the covering factor, for an evolving $d_{\rm abs}$ (Fig. \ref{abs_evolv}) and $d_{\rm QSO}$ (Fig. \ref{quasar_evolv}).}
\label{ratio-f}
\end{figure}

\begin{figure}
\centering \includegraphics[angle=-90,scale=0.48]{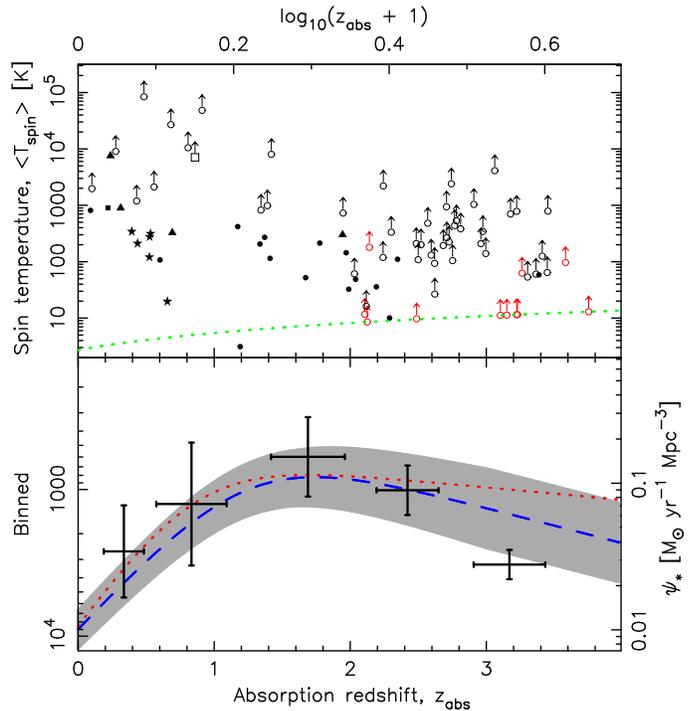}
\caption{As Fig. \ref{spin-bcw}, but including the statistical values for $d_{\rm abs}/d_{\rm QSO}$ and inverted, thus
yielding the statistical spin temperature. The dotted line in the top panel shows the lowest possible temperature,
due to the CMB at that redshift.
The scaling in the bottom panel suggests that  $\psi_{*}\approx100/T_{\rm spin}$ ~\Mo~yr$^{-1}$~Mpc$^{-3}$.}
\label{foverT}
\end{figure}

Applying the derived $d_{\rm abs}/d_{\rm QSO}$ (canonical) ratios to Equ.~\ref{f} we obtain the covering factors shown in Fig.~\ref{ratio-f} and 
from  
Equ. \ref{equ:NoverT} the {\em statistical spin temperature}, 
\[
\left<T_{\rm spin}\right>= \left\{   
\begin{array}{l l}
\frac{N_{\text{\HI}}}{1.823\times10^{18}}\left(\frac{d_{\rm abs} DA_{\rm QSO}}{d_{\rm QSO} DA_{\rm abs}}\right)^2 \frac{S_{\rm obs}}{\Delta S \Delta v} & \text{ if } \theta_{\rm abs} < \theta_{\rm QSO}\\
 & \\
\frac{N_{\text{\HI}}}{1.823\times10^{18}} \frac{S_{\rm obs}}{\Delta S \Delta v} & \text{ if } \theta_{\rm abs} \geq\theta_{\rm QSO},\\
\end{array}
\right.
\]
where $\theta_{\rm abs} = d_{\rm abs}/DA_{\rm abs}$ and $\theta_{\rm QSO} = d_{\rm QSO}/DA_{\rm QSO}$. Applying this to the data gives the distribution shown in Fig.~\ref{foverT},
from which we see the binned values of $\left<1/T_{\rm spin}\right>$ tighten around the evolution of $\psi_{*}$.  This
includes the errant high redshift bin, which now agrees with the star formation density of \citet{bwc13} to within
$1\sigma$.  Note that there is one point with $T_{\rm spin} = 3.1$ K, which is lower than the temperature of
the Cosmic Microwave Background (CMB) expected at $z_{\rm abs} = 1.191$ ($T_{\rm spin} = 6.0$ K, \citealt{mbb+12}). This
is unphysical, although it should be borne in mind that this is a statistical correction, with the binned values
occupying the same range as other studies (\citealt{lb01,kc02,sgp+12,rmgn13,kps+14,cras16}), where the covering factor
is estimated/assumed.

\subsection{Measured radio extents}
\label{others}

\begin{figure}
\centering 
\includegraphics[angle=270,scale=0.48]{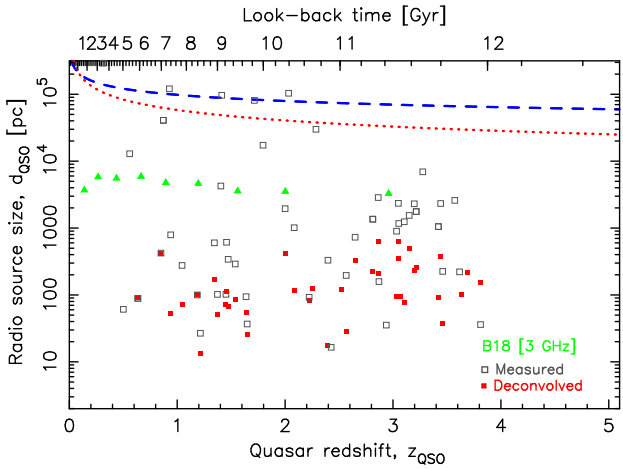}
\caption{As Fig. \ref{quasar_evolv}, but showing the evolution of $>40~\mu$Jy  AGN and star-forming galaxies at 3 GHz
(\citealt{bzc+18}, B08), in addition to the deconvolved (filled squares) and measured (unfilled squares) source 
sizes for the 21-cm searched DLAs.}
\label{quasar_log}
\end{figure}

\begin{figure}
\centering \includegraphics[angle=-90,scale=0.48]{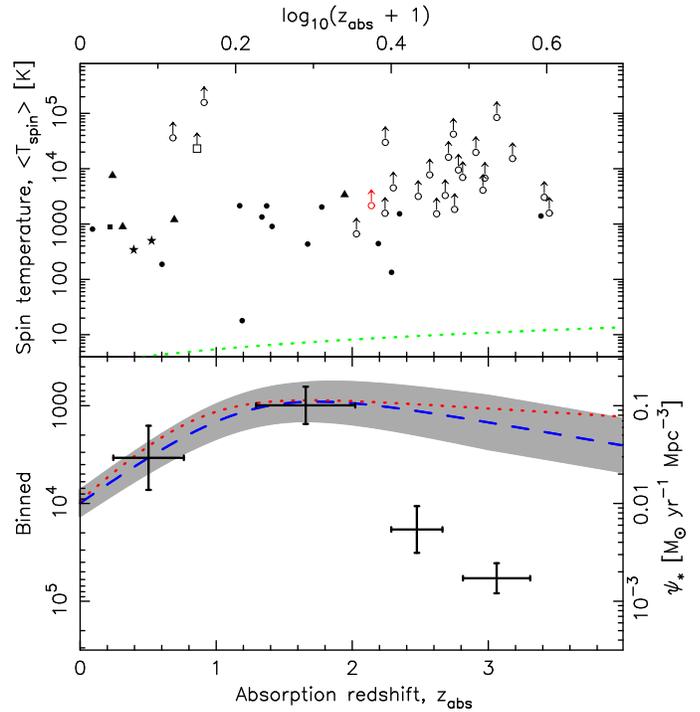}
\caption{As Fig. \ref{foverT}, but where the deconvolved source sizes are used to obtain $d_{\rm QSO}$.}
\label{T-decon}
\end{figure}

While several studies suggest that radio sources evolve from $\sim100 - 10$~kpc over $z=0 -
5$ \citep{nvkj93,sin93,wrbl99,ouoo18}, we now explore the possibility that the source size may be significantly smaller.
For example, for radio selected AGN and star-forming galaxies, \citet{bzc+18} find sizes of $\approx6 - 3$ kpc over $z=0
- 3$ for $\mu$Jy radio selected AGN and star-forming galaxies, although at 3~GHz. 
Furthermore, 49 of the background sources of the DLAs searched in 21-cm absorption have been 
imaged at high resolution \citep{klm+09,kem+12,kps+14,ekp+12} and found to have sizes of  $\approx0 - 700$ pc, which are 
significantly smaller than those obtained from the source size evolution (Fig. \ref{quasar_log}).
Applying the deconvolved sizes gives the statistical spin temperature distribution shown in Fig. \ref{T-decon},
where the covering factor is unity in all cases ($\theta_{\rm abs} > \theta_{\rm QSO}$), when using the 
evolving galaxy model (Fig.~\ref{abs_evolv}). While this gives a similar result to the canonical model
with  $\psi_{*}\approx100/T_{\rm spin}$ ~\Mo~yr$^{-1}$~Mpc$^{-3}$ (Fig.~\ref{foverT}) up to the
peak of $\psi_{*}$, we see a rapid rise in the spin temperature at $z_{\rm abs} \gapp2$ resulting in
a much more severe departure from the evolution of the star formation density.

\begin{figure}
\centering \includegraphics[angle=-90,scale=0.48]{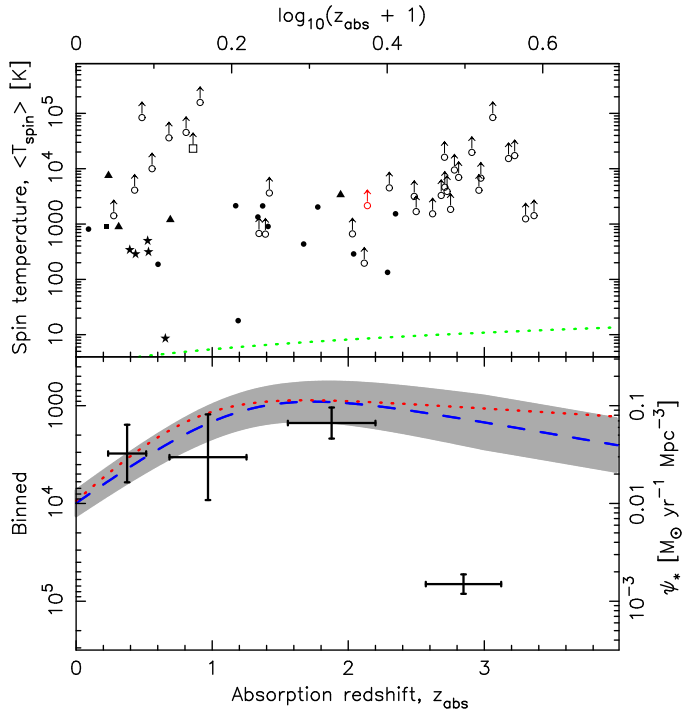}
\caption{As Fig. \ref{foverT}, but where the measured source sizes are used to obtain $d_{\rm QSO}$.}
\label{T-meas}
\end{figure}

These sizes are, however, obtained from the deconvolution of the radio images in which the emission is clearly resolved
in the convolved synthesised images, leading, in four cases, to unphysical  sizes of  $\approx0$ pc.
Therefore, in Fig.~\ref{quasar_log} we also show the measured source sizes as obtained directly
from the resolved radio images, in conjunction with others from the high resolution 
Very Large Baseline/Array (VLBI/VLA) observations.\footnote{Compiled from \citet{ujpf81,per82,skn82,huo83,gh84,sfwc84,au85,rpd87,bmsl88,bwl+89,nh90,sbom90,fpa92,vff+92,lbm93,mbp93,pgh+93,gsb+94,lgh94,pss+94,clc+95,pwx+95,rsa+95,bpf+96,cbr+96,fcf96,fc97,hsmr97,sob+97,swm+97,bwpw98,dbam98,shk+98,swm+98,tml+98,wbp+98,rkp99,bvl00,fc00,ffp+00,bgp+02,htt+07,gsp+12,sgp+12}.}
These give the statistical spin temperature distribution shown in  Fig.~\ref{T-meas}, 
which is similar to that of the deconvolved sizes, due to the fact that $f=1$ for the majority (48 out of 55) of the DLAs.

\begin{figure}
\centering 
\includegraphics[angle=270,scale=0.48]{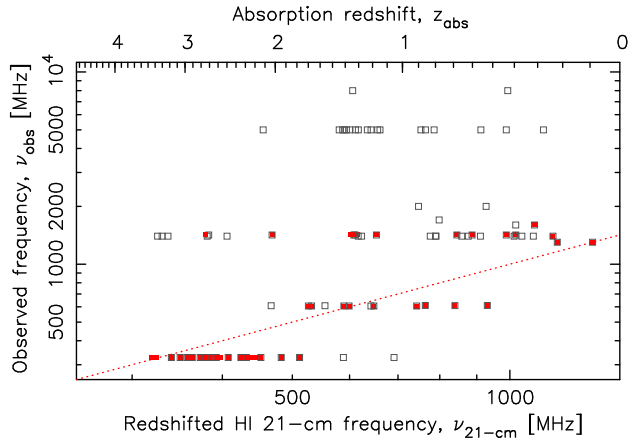}
\caption{The frequency at which the radio source size is determined versus the redshifted 21-cm frequency. The symbols
are as per Fig. \ref{quasar_log} and the line shows $\nu_{\rm obs} = \nu_{\rm 21-cm}$.}
\label{obs-21cm}
\end{figure}

The difference in the sizes between the general radio source ($\sim100$ kpc)
and DLA background  ($\sim10$ kpc) populations (Fig.\ref{quasar_log}) may be due to the high resolution radio 
images being obtained at high frequencies, where the radio source size will be smaller. This could be
 particularly acute for the 21-cm absorption, where $\nu_{\rm 21-cm}\ll1420$~MHz at high redshift. 
From Fig.~\ref{obs-21cm}, 
we see that this is often the case.
Furthermore,  high resolution imaging may resolve out all of the  extended flux, again underestimating the source size as
seen by the absorbing gas. 

\subsection{Physical implications}

\begin{figure}
\centering \includegraphics[angle=-90,scale=0.48]{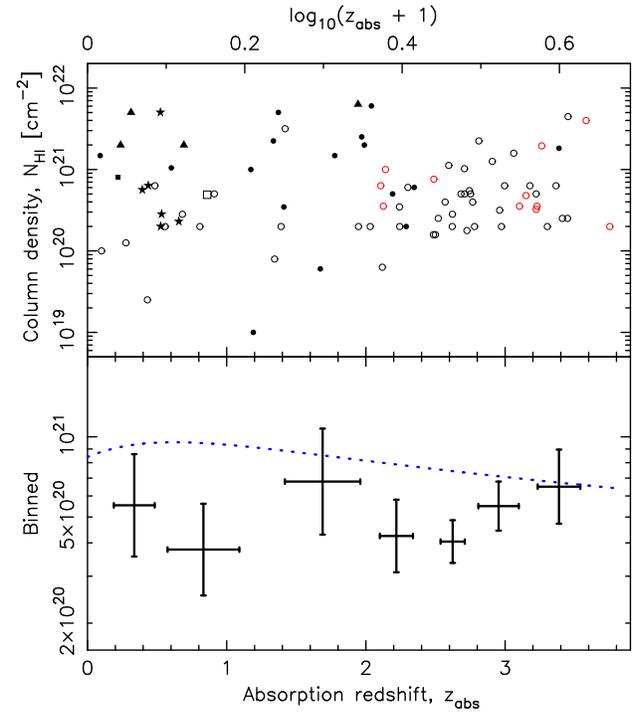}
\caption{The column density distribution of the DLAs searched in 21-cm absorption. 
A generalised non-parametric Kendall-tau test gives a probability of $P(\tau) = 0.82$ which is significant at just
$S(\tau) = 0.23\sigma$, assuming Gaussian statistics. That is, there is no correlation.
In the bottom panel, we show the binned values, with the dotted curve showing the mean column density obtained
from the best fit to the cosmological \HI\ mass density of \citet{cmp+17}, $\Omega_{\text{\HI}} = 4.0\times10^{-4}(z_{\rm abs}+1)^{0.60}$ (see \citealt{cur17a}). }
\label{N-z}
\end{figure}

For an evolving absorber size, the canonical model and deconvolved/measured radio extents predict very different spin
temperatures at high redshift, which may be evident through the degree of ionisation of the neutral gas.  For a given
number of baryons, an increase in the ionised gas, $N_{\text{\HII}}$, will be matched by a decrease in the neutral gas.
However, like the general DLA population, there is no evidence of an evolution in the mean column density (Fig. \ref{N-z}),
and so we do not expect any significant ionisation over and above that at lower redshift.\footnote{Unlike the atomic gas
  within the host galaxies of high redshift active sources, which is completely ionised at $\gapp10^{56}$ ionising
  photons s$^{-1}$ \citep{cw12,chj+19}.}

The Saha equation \citep{sah21,fri08} gives the ionisation fraction, the ratio of the number of ions,
${\cal N}_{\rm II}$, to the total number of ions plus atoms, ${\cal N}_{\rm I}$, 
at temperature $T$, as
\begin{equation}
\frac{x^2}{1-x} = \frac{1}{n}\left(\frac{2\pi m_{\rm e}kT}{h^2}\right)^{3/2}e^{-\phi/kT}, \text{ where } x\equiv \frac{{\cal N}_{\rm II}}{{\cal N}_{\rm I}+ {\cal N}_{\rm II}},
\label{saha}
\end{equation}
with $n$ being the number density of particles [m$^{-3}$], 
$m_{\rm e}$ the mass of the electron,  $k$ the Boltzmann
constant, $h$ the Planck constant and $\phi = 13.6$ eV the ionisation potential.

\begin{figure}
\centering \includegraphics[angle=-90,scale=0.48]{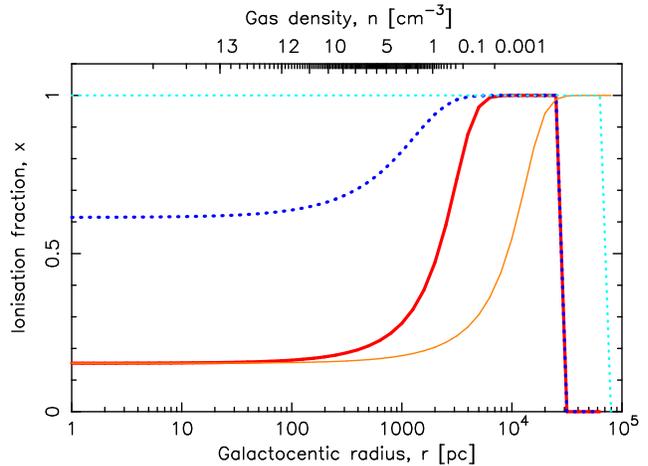}
\caption{The ionisation fraction versus the galactocentric radius at temperatures of 3250~K and (solid curves) 
3500~K (dotted lines) for a central gas density of $n_0 = 13.4$ \ccm. The thick traces are for a scale-length of $R=0.73$ kpc
and the thin traces for $R=3.15$ kpc.}
\label{ion-n}
\end{figure}

Assuming thermodynamic equilibrium across the disk and solving the quadratic in Equ. \ref{saha},
for an exponential gas disk, $n = n_0 e^{-r/R}$, we obtain the ionisation fraction profiles in Fig.~\ref{ion-n}.
These are shown for the gas distribution of the Milky Way ($n_0 = 13.4$ \ccm\ and $R_{\rm MW}=3.15$ kpc,  \citealt{kk09}),
as well as for an evolved absorber size, where  we may expect the scale-length of the 
atomic gas density to decrease as $R = R_{\rm MW} (z_{\rm abs}+1)^{-1.05}$, giving $R=0.73$ kpc at $z\sim3$ (Sect.~\ref{as}).
In either case, the gas completely is ionised at all radii for $T\gapp4000$~K, whereas 
at $T=3250$~K (the canonical values, Sect.~\ref{can}) the gas is mostly neutral out to radii of $\sim1$~kpc,
beyond which the density drops to below $n\sim1$~\ccm. 

Although similar to the canonical values at $z_{\rm abs}\lapp2$, the deconvolved/measured source sizes (Sect.~\ref{others}) in
conjunction with the evolved absorber sizes (Sect.~\ref{as}) give much higher statistical spin temperatures at $z_{\rm
  abs}\sim3$ ($T_{\rm spin}\approx60\,000$~K). This implies a very high ionisation fraction ($x=1$), to the point that
the strength of the Lyman-\AL\ absorption would be below that of Lyman Limit Systems ($N_{\text{\HI}}=10^{17.2} -
10^{20.3}$ \scm) at the level of the Lyman-\AL\ Forest, contrary to what is observed (Fig.~\ref{N-z}).

\begin{figure}
\centering \includegraphics[angle=-90,scale=0.48]{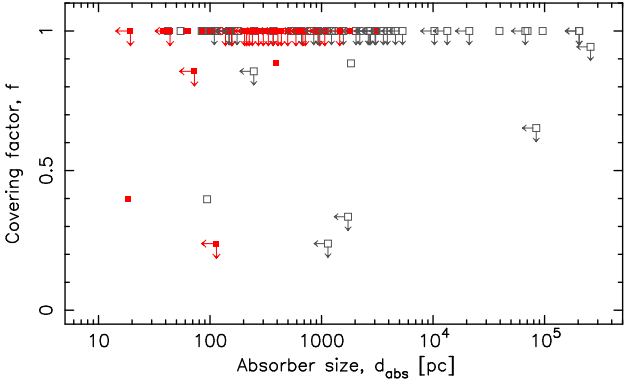}
\caption{The covering factors and absorber sizes required for a gas temperature of $T=3000$~K. The symbols are as per
Fig. \ref{quasar_log}, with the limits due to the 21-cm non-detections.}
\label{f-d}
\end{figure}

Dispensing with the evolved absorber sizes, we can use the fact that the gas is mostly neutral to 
estimate the covering factor required to ensure the gas is below this
temperature (Equ.~\ref{enew}), and thus the maximum permitted absorber extent  (Equ~\ref{f}).
From Fig.~\ref{f-d},
we see that most of the deconvolved source sizes and many of the measured sizes require absorber extents 
of $d_{\rm abs}\lapp1$~kpc. This is significantly smaller than
those predicted from the evolution of large galaxies, $d_{\rm abs} \approx10$ kpc at $z\sim3$ (Sect.~\ref{as}).
Taking the raw values (Fig.~\ref{spin-bcw}), 
the $z\sim3$ bin gives $T_{\rm spin}(d_{\rm QSO}/d_{\rm abs})^2\approx10\,000$~K, which compared with the 
$T\lapp4000$~K limit imposed by the ionisation fraction, gives $d_{\rm QSO}\gapp2 d_{\rm abs}$.

High redshift imaging of the \HI\ 21-cm emission from DLA hosts is required to verify that the \HI\ disk
extends as far past the optical emission at $z\gapp1$ as it does at lower redshift ($d_{\rm abs} \sim 5d_{\rm HL}$), 
although even the SKA will be limited to $z\lapp1$ in the detection of 21-cm spectral line emission \citep{so15}.
If the case, the above absorption cross-sections
imply $d_{\rm HL}\approx2$~kpc for the canonical model and $d_{\rm HL}\lapp200$ pc  for the measured radio source sizes.
DLAs are hypothesised to arise in a multitude of sources, including galactic disks, dwarf galaxies,
rapidly rotating proto-disks, merging sub-galactic systems and low surface brightness galaxies 
(\citealt{wtsc86,mmv97,pw97,hsr98,jbm99}), which does not constrain the above absorption cross-sections.
 Note, however, that  impact parameters of $b\gapp20$~kpc have been found at $z_{\rm abs}\gapp2$,
although identification of  DLA hosts is very rare at high redshift (see \citealt{fop+15} and references therein).   

\section{Summary}

We have used unpublished archival GMRT data of \HI\ 21-cm absorption searches in high redshift ($z_{\rm abs} \gapp2$)
damped Lyman-$\alpha$ absorption systems to determine whether the fraction of cold neutral gas continues to trace the
star formation density at these redshifts. This adds a dozen new data points, albeit limits, from which the spin
temperature, degenerate with the ratio of the absorber and the continuum source sizes, does increase at high
redshift. This is, however, higher than predicted by the star formation density [$T_{\rm spin} (d_{\rm QSO}/d_{\rm
  abs})^2\approx10\,000$ cf. $\approx2000$~K].  This could be explained by an evolution in $d_{\rm QSO}/d_{\rm abs}$ and
estimating this from the known evolution of galaxy and radio source sizes (the canonical model), gives $\left<T_{\rm
    spin}\right>\approx3000$~K, cf.  $\left<T_{\rm spin}\right>\approx2000$~K expected from the star formation density.
This may suggest that the UV data, from which $\psi_{*}$ is derived at high redshift, is still over-corrected for dust
obscuration (\citealt{bwc13}, cf. \citealt{hb06}). However, given that there is only a single 21-cm detection in this
bin, which forms the parent population with which to incorporate the limits, further $z_{\rm
  abs}\gapp3$ detections are required to verify this. Until then, the high redshift bin is best treated as a lower limit to 
$\left<T_{\rm spin}\right>$.

There exists high resolution radio imaging of many of the background QSOs, although applying the galaxy size evolution
yields $\left<T_{\rm spin}\right>\approx60\,000$~K at $z_{\rm abs}\sim3$, where all of the neutral gas would be ionised,
which is inconsistent with the observed flat evolution of the column density. We have used the limiting
$T\approx4000$~K, above which all of the gas is expected to be ionised, to constrain the required absorber extents
based upon the high resolution images. These require \HI\ extents of $d_{\rm abs}\lapp1$~kpc, or half-light radii of
$r_{\rm HL}\lapp100$~pc, which would support the hypothesis that high redshift DLAs arise predominately in dwarf
galaxies, although large impact parameters ($b\gapp20$~kpc) have been found at $z_{\rm
  abs}\gapp3$. 

The raw values [$T_{\rm spin} (d_{\rm QSO}/d_{\rm abs})^2$], which remove the bias between the low ($z_{\rm abs}\lapp1$)
and high redshift samples introduced by the geometry of an expanding Universe, do suggest that the spin
temperature, degenerate with the background source--absorber size ratio, traces the star formation
density, with the rogue $z_{\rm abs}\sim3$ bin being ``reigned in'' when canonical values of $d_{\rm abs}$ 
and $d_{\rm QSO}$ are applied. This suggests that
$\psi_{*}T_{\rm spin}\approx100$ ~\Mo~yr$^{-1}$~Mpc$^{-3}$~K, at least at $z_{\rm abs}\lapp3$.
Again, this may be due to the single detection in the parent population 
or could suggest over-corrections to the UV data, as well as a high redshift variation in the $d_{\rm QSO}/d_{\rm abs}$ ratio, perhaps
due to a differing ratio in the half-light/\HI\ radii.  This could be addressed with further 
\HI\ 21-cm detections in $z_{\rm abs}\gapp3$ DLAs, which, although difficult, is feasible with current instruments \citep{cur18}.

\section*{Acknowledgements}

I wish to thank the anonymous referee for their helpful comments.
This research has made use of the NASA/IPAC Extragalactic Database (NED) which is operated by the Jet Propulsion
Laboratory, California Institute of Technology, under contract with the National Aeronautics and Space Administration
and NASA's Astrophysics Data System Bibliographic Service.  This research has also made use of NASA's Astrophysics Data
System Bibliographic Service and {\sc asurv} Rev 1.2 \citep{lif92a}, which implements the methods presented in
\citet{ifn86}.


\label{lastpage}

\end{document}